\documentclass[twocolumn,secnumarabic,amssymb, nobibnotes, aps, prl]{revtex4}
\usepackage{amsmath}
 \usepackage{amsmath}
    \usepackage{amssymb}
    \usepackage{amsthm}
    \usepackage{amsfonts}
    \usepackage{braket}

\usepackage{color}
\usepackage{pstricks}
\usepackage{graphicx}
\usepackage{slashed}

\begin{document}

\title{Edge Modes and  Teleportation in a Topologically Insulating Quantum Wire}

\author{Majd Ghrear, Brie Mackovic,  Gordon W. Semenoff} 
\affiliation{ 
 Department of Physics and
Astronomy, University of British Columbia, 6224 Agricultural Road,
Vancouver, BC Canada V6T 1Z1 }

\begin{abstract}
  We find a simple model of an insulating state of a quantum wire which has a single  isolated edge mode.  We argue that, when brought to proximity, the edge modes  on independent wires naturally form Bell entangled states 
 which could  be used for elementary quantum processes such as teleportation. We give an example of an algorithm which teleports the spin state of an electron from one quantum wire to another.  
   \end{abstract}

\maketitle

   An important step in the evolution of quantum circuit technology would be the ability to reliably 
create Einstein-Podolsky-Rosen (EPR) entangled 
electron pairs at controllable locations in an electronic system. This would
raise the possibility of making elementary quantum devices  using the many existing technologies available for the fabrication and manipulation of  electronic materials \cite{epr1}-\cite{epr3}.
An  obstacle to the use of electrons for EPR pairs is their susceptibility to decoherence which results from  
their interactions with their environment and with each other.   Some effort has gone into inventing  systems which minimize this decoherence by,  for example,  dynamically isolating the electron state in a band gap, minimizing its degeneracy and screening its charge.  Kitaev's famous proposal of ``Majorana fermions'' is a good example \cite{Kitaev:2001kla} which achieves all of these and where possibilities of interesting non-local phenomena have been explored \cite{Semenoff:2006zp}-\cite{Ren:2017ytv}. Braiding of Majorana fermion histories is one of the prime candidates for quantum computation  \cite{Kitaev:1997wr}-\cite{Litinski:2018pgo}. However, a single Majorana state as a qubit has the drawback that it is a superposition of two states which differ in fermion number parity, $(-1)^F$, and the states where the Majorana level is occupied and unoccupied cannot be mixed with physical quantum gates. For a system with a few qubits, it can be useful to have more degeneracy. 

 In this letter, we shall seek an alternative using ordinary electrons which have both charge and spin. We isolate the states dynamically by considering modes at the center of a band gap.  We shall propose a dynamical mechanism for combating decoherence.   We will design a system where the EPR state is dynamically favoured in that it is the ground state of a subsystem, so that, even if it decoheres by jumping to an excited state, its natural dynamical evolution would let it decay to the ground state and thereby restore the coherence.  As well as minimizing decoherence,  this approach   yields a tool for processing the system. Beyond applying quantum gates and performing quantum measurements, we can allow the system to go through a possibly dissipative relaxation from an excited state to its ground state. We will then use this additional tool in an essential way, to design an algorithm where the EPR state is used as a resource to teleport the spin wave-function of an electron between quantum wires. Although the material  and the gates needed for the process are hypothetical, we conform to certain rules which are aimed at utility.  We restrict ourselves to one and two-qubit gates which act on electron spins only, and, in particular, they are local operations and they respect electric charge and fermion number parity superselection rules. 

\begin{figure}
\includegraphics[scale=.3]{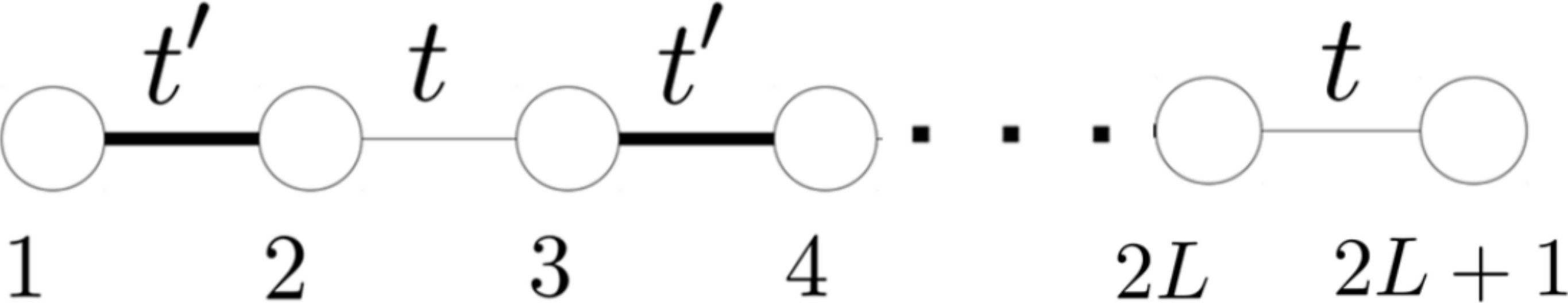}\\
\begin{caption} {  The odd-sited one-dimensional lattice with  alternating long $t$ and short $t'$
bonds.      \label{polyacetylene}  
}\end{caption}
 \end{figure}
\begin{figure}
\includegraphics[scale=.4]{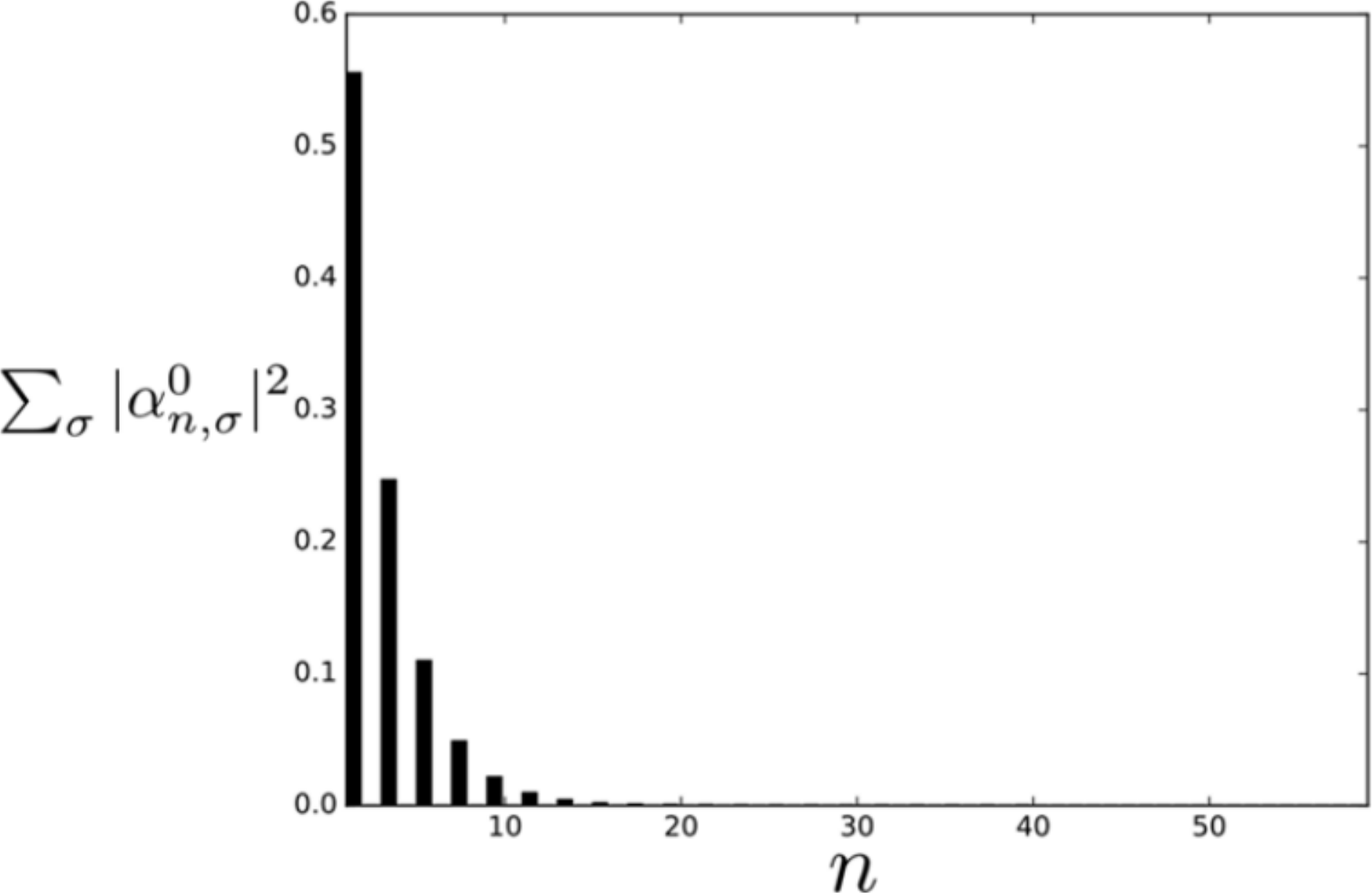}\\
\begin{caption} {     The probability density of the zero mode, $\sum_\sigma \left|\alpha_{n,\sigma}^0\right|^2$ with $2L+1=59$ and  
  $t'/t=2/3$.      \label{figure2}  
}\end{caption}
 \end{figure}

 
 We will make use of the fact that, in a system of noninteracting fermions, where there are single-particle levels with both positive and negative
 energies, and when there is a unitary transformation which maps positive energy levels onto negative energy levels (particle-hole symmetry), so that positive and negative single-particle states are paired, and when the number of energy levels must be odd, there is necessarily an unpaired zero energy state.   Then, it is a simple exercise to design a system where that zero energy state is isolated in a band gap and it is an edge state.  
Our model is a system of spin-$\frac{1}{2}$ fermions bound to the atomic sites of a 
 one-dimensional array of atoms. The lattice sites are considered immobile and
 the fermions are  described using the tight-binding approximation.                                                                                                                                                                                                                                                                                                                                                                                                                                                                                                                                                                                                                                                                                                                                                                          
The array has an odd number of sites, labeled $n=1,2,\ldots,2L+1$,  
 and $\alpha^\dagger_{n,\sigma}$ and $\alpha_{n,\sigma}$ 
 are the creation and annihilation operators for aa fermion occupying site $n$ with spin $\sigma=\uparrow,\downarrow$.
 They obey  the anti-commutator algebra
  \begin{align}
  \left\{\alpha_{n,\sigma},\alpha^\dagger_{n',\sigma'}\right\}&=\delta_{nn'}\delta_{\sigma\sigma'}
  \\
  \left\{\alpha_{n,\sigma},\alpha_{n',\sigma'}\right\}&=0=\left\{\alpha_{n,\sigma}^\dagger,\alpha_{n',\sigma'}^\dagger\right\}
  \end{align}
The Hamiltonian  describes spin-independent tunnelling between nearest neighbours 
   \begin{align}\label{Hamiltonian}
H=\sum_{\sigma=\uparrow,\downarrow}\sum_{m=1}^L \left( t^*\alpha_{2m ,\sigma}^\dagger\alpha_{2m+1 ,\sigma}+ t\alpha_{2m+1 ,\sigma}^\dagger\alpha_{2m ,\sigma}\right)
\nonumber \\+\sum_{\sigma=\uparrow,\downarrow}\sum_{m=1}^{L}  \left({t'}^*\alpha_{2m ,\sigma}^\dagger\alpha_{2m-1 ,\sigma}+{t'}\alpha_{2m-1 ,\sigma}^\dagger\alpha_{2m ,\sigma}\right)
 \end{align} 
 We shall  assume that the strengths of the tunnelling amplitude $t$ for moving to the right from an even site and $t'$ for moving to the left from an even site are different.  These alternating ``long'' and ``short'' bonds,  depicted in Fig.\ \ref{polyacetylene},  are similar to those in a polyacetylene molecule \cite{Niemi:1984vz}.    We shall assume that the phases of the creation and annihilation operators are adjusted so that $t $ and $t'$ are real and positive.  
 It is important that the total number of sites, $2L+1$,  is odd.   The Schr\"odinger equation 
 \begin{align}
i\hbar\frac{d}{d\tau} \alpha_{2m,\sigma}(\tau)= t\alpha_{2m+1,\sigma}(\tau) +t'\alpha_{2m-1,\sigma} (\tau)\\
i\hbar\frac{d}{d\tau}  \alpha_{2m+1,\sigma}(\tau)  = t'\alpha_{2m+2,\sigma}(\tau) +t\alpha_{2m,\sigma} (\tau)
  \end{align}
  with boundary conditions $\alpha_{ 0,\sigma}=0=\alpha_{ 2L+2,\sigma}$ 
is solved by the  stationary states, 
\begin{align}
&\alpha_{2m,\sigma}(\tau)=\tfrac{e^{-i\varepsilon \tau/\hbar}}{\sqrt{L+1}}\sin\left(\tfrac{\pi k}{L+1}m\right)~,~k=1,2,\ldots, L
\nonumber
\\   
&  \alpha_{2m+1,\sigma}(\tau)= \tfrac{e^{-i\varepsilon 
\tau/\hbar}}{\varepsilon\sqrt{L+1}}\left[  t'  \sin\left(\tfrac{\pi k(m+1)}{L+1}\right)+t
\sin\left(\tfrac{\pi km}{L+1}\right)  \right] 
\nonumber
\end{align}
with positive (conduction) and negative (valence) energy bands   and   band gap   $2|t-t'|$, 
 \begin{align}
  \varepsilon(k)=\pm\sqrt{ (t-t')^2+4tt'\cos^2\left[\frac{\pi k}{2L+2}\right] }
  \end{align}
  The spectrum is symmetric. For each $k$ there is a positive and negative energy $\varepsilon(k)$.   
 In addition, there is a single mid-gap zero mode, with $\varepsilon=0$, and wave-function
  \begin{align}\label{zeromode}
  \alpha_{2n,\sigma}^0(\tau)=  0,~
\alpha_{2n+1,\sigma}^0(\tau)=\sqrt{\frac{1-(t'/t)^2}{1-(t'/t)^{2L+2}}} \left(-\frac{t'}{t}\right)^n 
  \end{align}
 which is concentrated near $n=0$ if $t'<t$ and near $n=L$ if $t'>t$. The probability density
   $\sum_\sigma \left|\alpha_{n,\sigma}^0\right|^2$   with $2L+1=59$ and 
  $t'/t=2/3$ is depicted in Fig.\ \ref{figure2}.  The existence of the energy gap and the mid-gap zero mode
 concentrated at one end of the wire will be important to us.  The number of sites is not important as long as this number is odd, in fact the wire could be
 very short if that is convenient. 
 
  The existence of the zero mode is not an accident.   As we have discussed, the spectrum is symmetric about $\varepsilon=0$.
  If $\alpha_{n,\sigma}$ is a solution of the Schr\"odinger equation with energy $\varepsilon$, then $(-1)^n\alpha_{n,\sigma}$ is
  also a solution with energy $-\varepsilon$.  The non-zero energy states thus come in positive-negative pairs. 
  Moreover, there must be an odd number
  of states, so there must necessarily be an un-paired zero mode.  This feature is ``topological'' in the sense that it would be
  preserved by any modification of the Schr\"odinger equation which respected the particle-hole symmetry and the odd number of states.

 \begin{figure}
\includegraphics[scale=.3]{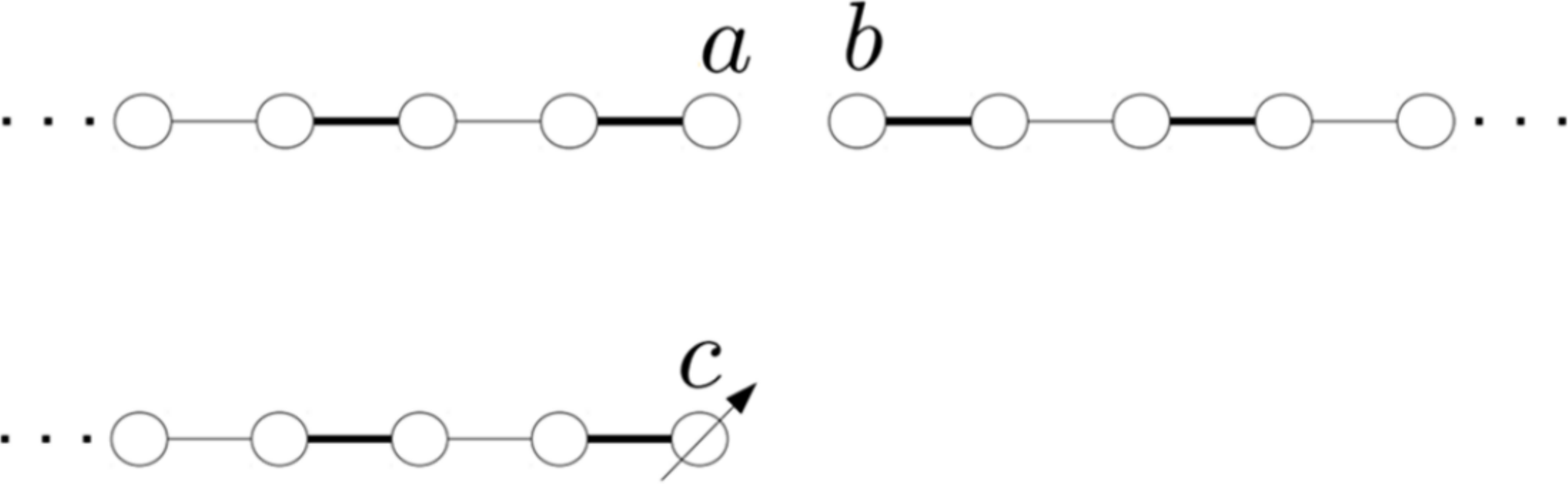}\\
\begin{caption} {  The $a$, $b$ and $c$ systems as edge states.  $c$ is occupied by an electron in a particular spin state and $a$ and $b$ are brought into proximity.  The spin wave-function of the electron occupying $c$ is to be teleported to a single electron occupying $b$.   The $c$-electron does not interact with $a$ or $b$.  $a$ and $b$ interact via $H_{\rm int}$ in (\ref{int}) which favours a state with $a$ and $b$ both singly occupied and together forming a spin singlet.   Alice applies one and two qubit gates to $a$ and $c$ and Bob can apply gates to $b$.    \label{figure3}  
}\end{caption}
 \end{figure}  
    
   In the following discussion, we shall consider the many-electron state which has all of the negative energy modes occupied and all of the positive energy modes empty.    We assume that 
  the energy gap  is sufficiently large compared to the energy scales  of phenomena of interest that
  the states with $|\varepsilon|\geq |t-t'|$ decouple and play no role.  The active mode is the zero mode with wave-function (\ref{zeromode})  concentrated at one of the
   edges of the wire.  Electrons occupying the zero mode are annihilated and  created   by $(a_\sigma, a_{\sigma}^\dagger)$ with the algebra
   \begin{align}
   \left\{ a_\sigma, a^\dagger_{\sigma'}\right\} = \delta_{\sigma\sigma'}~,~
    \left\{ a_\sigma, a_{\sigma'}\right\} =0= \left\{ a^\dagger_\sigma, a^\dagger_{\sigma'}\right\} 
    \end{align}
     The space of states where the spin up or  down
    zero modes are populated or unpopulated form a two qubit system with a four-dimensional Hilbert space, spanned by
    $$
    \{ |0>,a_\uparrow^\dagger|0>,a_\downarrow^\dagger|0>,a_\uparrow^\dagger a_\downarrow^\dagger|0>\}
    $$ 
    where $|0>$ is the state with the zero modes (and all other zero mode electron states which we shall discuss in the following)
    empty, $a_\uparrow |0>=0$ and $a_\downarrow |0>=0$. The ground state of the charge neutral system is  two-fold degenerate in the span of 
    $\{a_\uparrow^\dagger |0>,a_\downarrow^\dagger |0>\}$.  It has charge zero and spin $\frac{1}{2}$.  The ``particle'' and ``hole'' states $a_\uparrow^\dagger a_\downarrow^\dagger |0>$ and $|0>$ have charges $+1$ and $-1$, respectively, 
    and they are spin singlets.  These unusual spin-charge relations, which have long been associated with fractionally charged solitons in polyacetylene \cite{Niemi:1984vz} \cite{Su:1979ua},
 in this case, are a simple consequence that the lattice has an odd number of sites. 
    
    Consider the scenario where edges containing zero modes of two wires are brought into proximity, depicted in
   Fig.\ \ref{figure3}. 
   We will describe the zero mode states of one wire as above, and the other similarly, with 
   $(a_\sigma,a_\sigma^\dagger)$ replaced by 
   $(b_\sigma,b_\sigma^\dagger)$. 
We model the interaction of the edge modes of the two wires by the Hubbard Hamiltonian
\begin{align}\label{int}
&H_{\rm int}=\frac{e^2}{2}\left(\sum_{\sigma}a^\dagger_\sigma a_\sigma-1\right)^2+\frac{e^2}{2}\left(\sum_{\sigma}b^\dagger_\sigma b_\sigma-1\right)^2+H_\lambda
 \\
&H_{\lambda}=\sum_{\sigma=\uparrow,\downarrow}\lambda \left(a^\dagger_\sigma b_\sigma+b^\dagger_\sigma a_\sigma\right)
\label{lambda}\end{align}
where $\lambda$ is real and positive and $ e^2>>\lambda$.   The first two terms  in $H_{\rm int}$ model the Coulomb energy which discourages 
nonzero charge at a site (the -1 is the charge of the ion at each lattice site). The first two
terms in $H_{\rm int}$  have a degenerate, zero energy ground state with basis states $\{a^\dagger_\uparrow b_\uparrow^\dagger|0>, a^\dagger_\uparrow b_\downarrow^\dagger|0>,a^\dagger_\downarrow b_\uparrow^\dagger|0>,a^\dagger_\downarrow b_\downarrow^\dagger|0>\}$.  The third term in (\ref{int}), $H_\lambda$ in (\ref{lambda}), 
allows the electrons to hop between the edge states of the two wires. This weaker interaction splits the degeneracy of the ground state
of the first two terms so that the lowest energy state is the spin singlet, 
\begin{align}
\label{g}
|g>=\frac{1}{\sqrt{2}}\left[ a^\dagger_\uparrow b_\downarrow^\dagger+ b_\uparrow^\dagger a^\dagger_\downarrow\right]|0>
\end{align} 
and the excited state is a spin triplet.  The energy of the ground state, $|g>$, in second order perturbation theory, is $E_0=-4\lambda^2/e^2$
and the spin triplet has energy $E_1=0+{\mathcal O}(\lambda^4/e^6)$.    We shall assume that
 $\lambda^2/e^2<<|t-t'|$ so that these states remain well within the energy gap. 
We assume that the hopping Hamiltonian $ H_{\lambda}$ can be turned on and off at will.   

Then, on top of this, 
we shall  assume that there is an additional, controllable, dissipative interaction which stimulates  
transitions between the energy levels of $H_{\rm int}$.  We need this dissipation so that the quantum state relaxes
to the ground state, $|g>$,  of $H_{\rm int}$ discussed above.   
We will also need the  assumption that, although it dissipates energy, the dissipative process conserves the electric charge and the 
spin of the $a$-$b$ system.   

The state $|g>$ in (\ref{g})  is an entangled state of the $a$-$b$ system.  We will use it as a resource for quantum teleportation \cite{telep}. 
We shall assume that Alice has access to the $a$-system and Bob has access to the $b$-system. 
Now,  we assume that Alice is in possession of another electron 
 which is in a particular spin state, expressed as a superposition of its spin up and spin down
 states.  It is dynamically independent of the $a$-$b$ system, so that the total wave-function is the product state
 \begin{align}\label{psi}
 \psi= \frac{  g_1c_\uparrow^\dagger+g_2 c_\downarrow^\dagger }{\sqrt{2 }} \left[ a^\dagger_\uparrow b_\downarrow^\dagger+ b_\uparrow^\dagger a^\dagger_\downarrow\right]|0>
\end{align}
where $|g_1|^2+|g_2|^2=1$. 
Alice has access to the composite $a$-$c$ system. Her goal is to teleport the spin wave-function of the $c$-electron
to Bob.  Alice should now turn off $H_{\lambda}$ and do a measurement in a Bell basis of the $c$-$a$ system.  Equivalently,
she can perform the following three steps:
 Alice applies the two-qubit CNOT gate to the  $a$-$c$ system.  
This gate leaves the terms with $c_\uparrow^\dagger $ unchanged and it interchanges $a^\dagger_\uparrow$ and $a^\dagger_\downarrow$ in the terms with $c^\dagger_\downarrow$ . The state vector becomes
\begin{equation}
\psi=  \left( \frac{g_1}{\sqrt[]{2}} c_{\uparrow}^\dagger (a_{\uparrow}^\dagger b_{\downarrow}^\dagger +  b_{\uparrow}^\dagger a_{\downarrow}^\dagger ) + \frac{g_2}{\sqrt[]{2}} c_{\downarrow}^\dagger (a_{\downarrow}^\dagger b_{\downarrow}^\dagger +   b_{\uparrow}^\dagger a_{\uparrow}^\dagger) \right)  \ket{0}.
\end{equation}
Then, Alice applies the one-qubit Hadamard gate to the $c$-spins.  This gate replaces $c_\uparrow^\dagger$ by 
$\frac{c_\uparrow^\dagger+c_\downarrow^\dagger }{\sqrt{2}}$ and $c_\downarrow^\dagger$ by
$\frac{c_\uparrow^\dagger-c_\downarrow^\dagger}{\sqrt{2}}$ 
 to put the state vectors in the form
\begin{align}
\psi=  \left( \frac{g_1}{\sqrt[]{2}}  \frac{c_{\uparrow}^\dagger + c_{\downarrow}^\dagger}{\sqrt[]{2}} ( {a}_{\uparrow}^\dagger b_{\downarrow}^\dagger + b_{\uparrow}^\dagger a_{\downarrow}^\dagger ) + \frac{g_2}{\sqrt[]{2}}  \frac{c_{\uparrow}^\dagger - c_{\downarrow}^\dagger}{\sqrt[]{2}} ( {a}_{\downarrow}^\dagger b_{\downarrow}^\dagger +  b_{\uparrow}^\dagger a_{\uparrow}^\dagger) \right)  \ket{0}.
 \nonumber
\end{align}
Next, Alice measures the total spin, $J$, and the $z$-component of the spin, $J_z$, of her system.  The results for $J$, $J_z$ and the 
resulting state-vectors are
\begin{align}
&J=1, J_z=1:~~~~~
 c_{\uparrow}^\dagger a_{\uparrow}^\dagger  (g_1 b_{\downarrow}^\dagger - g_2b_{\uparrow}^\dagger)   \ket{0}
\\
&J=1, J_z=0:~~~~~\nonumber \\ & 
~~~~~\frac{c_{\uparrow}^\dagger a_{\downarrow}^\dagger+c_{\downarrow}^\dagger a_{\uparrow}^\dagger}{\sqrt[]{2}} \left(g_1  \frac{ {b}_{\uparrow}^\dagger -b_{\downarrow}^\dagger}{\sqrt[]{2}} - g_2 \frac{ {b}_{\uparrow}^\dagger +b_{\downarrow}^\dagger}{\sqrt[]{2}}\right)   \ket{0}
\\
&J=1, J_z=-1:~~~~~
 c_{\downarrow}^\dagger a_{\downarrow}^\dagger  (g_1 b_{\uparrow}^\dagger + g_2b_{\downarrow}^\dagger)   \ket{0}
\\
&J=0, J_z=0:~~~~~\nonumber \\ & 
~~~~~\frac{c_{\uparrow}^\dagger a_{\downarrow}^\dagger-c_{\downarrow}^\dagger a_{\uparrow}^\dagger}{\sqrt[]{2}} 
\left(g_1  \frac{ {b}_{\uparrow}^\dagger +b_{\downarrow}^\dagger}{\sqrt[]{2}} + g_2 \frac{ {b}_{\uparrow}^\dagger -b_{\downarrow}^\dagger}{\sqrt[]{2}}\right)   \ket{0}
\end{align}
Alice then uses a classical communication channel to communicate the two bits of data, $J,J_z$,  to Bob.  This tells Bob which gates to apply to recover
the original wave-function:
\[ \begin{array}{lcr}
\mbox{J} & J_z  & {\rm Bob~ gate} \\
\mbox{1} &1& iY\\
\mbox{1} &0 & ~~~H~{\rm  then}~ iY\\
\mbox{1} & -1 &I
\\
\mbox{0} & 0 & H \end{array}\] 
where $H$ is the Hadamard gate and $iY (b_\uparrow,b_\downarrow)=(b_\downarrow,-b_\uparrow)$.   
As a result, Bob has placed the $b$-electron into the same spin state that the $c$-electron originally occupied.  The spin state has been teleported. 

Alternative to an electronic system, one might envisage engineering such a system with spin $\frac{1}{2}$ fermion atoms occupying an optical lattice.  
In that case, the atoms
would have only a weak repulsion, the parameter $e^2$ would be small and the interaction between the zero modes on wires in 
proximity would be dominated by $H_\lambda$. In that case,
the ground state of $H_\lambda$ is $\frac{1}{2}(a^\dagger_\uparrow-b^\dagger_\uparrow)(a_\downarrow^\dagger-b_\downarrow^\dagger)|0>$  
which is a spin singlet, like (\ref{g}), but it contains  components where the edge modes are doubly occupied.   Alice then begins with
the state
\begin{align}\label{psi''}
 \frac{1}{2 }\left( g_1c_\uparrow^\dagger+g_2 c_\downarrow^\dagger\right) (a^\dagger_\uparrow-b^\dagger_\uparrow)(a_\downarrow^\dagger-b_\downarrow^\dagger)|0>
\end{align}
She should
first do a measurement to see whether her $c$-$a$ system is in a state of integer or half-odd-integer spin.  If it is integer spin, the measurement
projects onto the state in (\ref{psi}) and she can proceed with the teleportation protocol as before.   If the result is half-odd-integer, the 
state thereafter is 
  \begin{align} \label{1/2}
 \psi' =\frac{1}{\sqrt{2} }\left( g_1c_\uparrow^\dagger+g_2 c_\downarrow^\dagger\right) \left[
a_\uparrow^\dagger a_\downarrow^\dagger+b_\uparrow^\dagger b_\downarrow^\dagger
\right]|0>
\end{align}
We have not devised a simple algorithm for teleporting the spin wave-function to Bob when this is the state.  The most efficient way to proceed is to simply start again,
to turn on $H_{\lambda}$ and dissipation and to allow the $a$-$b$ subsystem to relax to
its ground state, 
$$
 \frac{1}{\sqrt{2} }\left( g_1c_\uparrow^\dagger+g_2 c_\downarrow^\dagger\right) \left[
a_\uparrow^\dagger a_\downarrow^\dagger+b_\uparrow^\dagger b_\downarrow^\dagger
\right]|0>
\to~~~~~~~~~~
$$
$$
~~~~~  \frac{1}{2 }\left( g_1c_\uparrow^\dagger+g_2 c_\downarrow^\dagger\right) (a^\dagger_\uparrow-b^\dagger_\uparrow)(a_\downarrow^\dagger-b_\downarrow^\dagger)|0>
$$
This evolution conserves spin and charge, there is only energy dissipation. It restores the state 
(\ref{psi''}). 
Then, Alice once again measures the total spin.   If it is integer, she turns off $H_\lambda$ and proceeds with the teleportation protocol.  If it is half-odd-integer, 
she repeats the restart.  Each time, the probability of obtaining integer spin is $\frac{1}{2}$ so that   the probability of failure after $n$ tries is $\frac{1}{2^n}$ which converges to zero. By this process, Alice  succeeds in teleporting the spin wave-function,  from the test electron to which Alice has access,  to the edge electron on Bob's quantum wire.   

 The most important role of $H_\lambda$ and dissipation in the above discussion  can be seen as the production of a charge neutral
 spinless state of the $a$-$b$ system. In fact, we could be more general and assume that the resource $a$-$b$ system is in a mixed state composed of all of
 its states which are charge neutral and have spin zero, that is, that   the state is described by any  density matrix in the vector space spanned by
 $\{ a_\uparrow^\dagger a_\downarrow^\dagger|0>,b_\uparrow^\dagger b_\downarrow^\dagger|0>,\frac{1}{\sqrt 2} (a^\dagger_\uparrow b_\downarrow^\dagger + b^\dagger_\uparrow a_\downarrow^\dagger )|0>\}$ which has a non-zero diagonal component in the direction $\frac{1}{\sqrt 2}(a^\dagger_\uparrow b_\downarrow^\dagger + b^\dagger_\uparrow a_\downarrow^\dagger )|0>$.  Again, when Alice measures the total spin of her subsystem, she either obtains integer spin, with the
pure state (\ref{psi})  whence
she can complete the teleportation, or a mixed state made in the space spanned by $$ \left( g_1c_\uparrow^\dagger+g_2 c_\downarrow^\dagger\right)a_\uparrow^\dagger a_\downarrow^\dagger|0>~,~\left( g_1c_\uparrow^\dagger+g_2 c_\downarrow^\dagger\right)b_\uparrow^\dagger b_\downarrow^\dagger|0>$$ whence she should redo the process.

The  zero modes associated with solitons in polyacetylene \cite{Su:1979ua} \cite{Su:1979wc} were originally 
discussed in the context  where the associated  mid-gap states lead to exotic quantum numbers such as fractional fermion number \cite{Niemi:1984vz}  \cite{Jackiw:1981wc}.  The disentanglement of the states of paired zero modes in order to find localized fractional charges  dates to that time \cite{Jackiw:1983uf}. 
The polyacetylene model, with alternating strong and weak bonds has since become one of the primary examples of a one-dimensional topological insulator \cite{top}, although to our knowledge the case where the number of lattice sites is odd and there is only one mid-gap state, residing on one of
the two edges of the wire has not been considered before.  It would be interesting to study this scenario with proximity induced superconductivity to understand the fate of the Majorana mode which one might expect to emerge there. 
 The polyacetylene molecule  could be a candidate for the insulator that we are discussing, if its chain can contain an odd number of sites.  
  An alternative would be to implement such a system using cold atoms in an optical lattice \cite{Ruostekoski:2007yp}
  \cite{ol}.

\noindent
This work was supported in part by NSERC.

\end{document}